\documentstyle[prl,aps,epsfig,twocolumn]{revtex}

\begin{document} 
\preprint{submitted to Phys. Rev. B}

\twocolumn[\hsize\textwidth\columnwidth\hsize\csname@twocolumnfalse\endcsname
\draft 

\title{
Coherent superposition of photon- and phonon-assisted tunneling \\ 
in coupled quantum dots
}

\draft
\author{
H. Qin$^{*}$, A.W. Holleitner$^{*}$, K. Eberl$^{\dagger}$, and R.H. Blick$^{*}$
}

\address{
$^{*}$~Center for NanoScience, Ludwig-Maximilians-Universit\"at, \\
Geschwister-Scholl-Platz 1, 80539 M\"unchen, Germany. \\
$^{\dagger}$~Max-Planck-Institut f\"ur Festk\"orperforschung, \\
Heisenbergstr. 1, 70569 Stuttgart, Germany.
}

\date{\today}
\widetext
\maketitle

\begin{abstract}
We report on electron transport through an artificial molecule 
formed by two tunnel coupled quantum dots, which are laterally confined 
in a two-dimensional electron system of an Al$_x$Ga$_{1-x}$As/GaAs heterostructure.
Coherent molecular states in the coupled dots are probed by photon-assisted tunneling (PAT).   
Above 10~GHz, we observe clear PAT as a result of 
the resonance between the microwave photons and the molecular states. 
Below 8~GHz, a pronounced superposition of phonon- and photon-assisted 
tunneling is observed. 
Coherent superposition of molecular states persists 
under excitation of acoustic phonons. 
\end{abstract}

]

\narrowtext
An artificial molecule is defined by coupling two quantum dots, leading to the formation of coherent 
electronic states in this molecule 
~\cite{kouwenhoven-book97,livermore-science96,blick-prl98,oosterkamp-nature98,stafford-prl96,stoof-prb96,holleitner-condmat00}. 
As a controllable two-level system, double dots are proposed for realizing a single quantum bit 
in solid state systems~\cite{loss-pra98}. 
One of the obstacles for such an application is the considerable dephasing induced 
by the semiconductor host materials. 
The investigation of dissipation processes in these nanostructures is thus of great importance. 
Operating such a double dot in the regime of weak coupling to the contacts ($\Gamma_{l,r} \rightarrow 0$) 
minimizes the dephasing of molecular states via the electron-electron interaction and it is then possible 
to study the interaction between confined molecular states and discrete phonon modes within the dots. 
Recently, Fujisawa~{\it et al.}~\cite{fujisawa-science98} found that at low temperatures vacuum fluctuations 
indeed result in pronounced transitions between energetically well-separated quantum states, 
generating discrete phonon modes. 
A theoretical treatment of this effect was then given by Brandes and Kramer~\cite{brandes-prl99}. 
However, the limitation of this experiment is that it can only be applied at large bias 
at which coherent molecular states can not be probed. 

Here, we present transport measurements on a double-dot molecule coupled to 
both microwave photons \emph{and} acoustic phonons. 
An external microwave source generates photons and piezoelectrically induces coherent acoustic phonons 
in phonon cavities formed by the Schottky gates.  
At near-zero bias, the formation of molecular states is revealed through 
single- and double-photon assisted tunneling (PAT). 
In addition, the combination of PAT with acoustic phonon-assisted tunneling 
is observed. We verify the persistence of coherent molecular states 
under phonon excitation. 

\begin{figure}[!h]
\begin{center}
\epsfig{figure=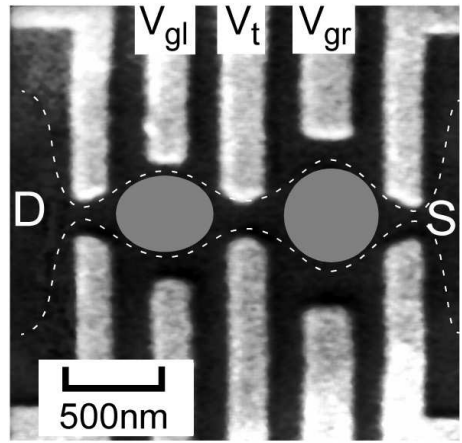,width=6cm,keepaspectratio}
\caption{Scanning electron microscope graph of this double dot. 
The dashed lines schematically show the 
edge of 2DEG. Two tunnel-coupled quantum dots are formed. 
} \label{structure}
\end{center}
\end{figure}

As shown in Fig.~\ref{structure}, five pairs of Schottky gates are defined by 
electron beam lithography and evaporation of gold 
on the surface of an Al$_x$Ga$_{1-x}$As/GaAs heterostructure. 
By applying proper negative voltages to these gates 
two quantum dots are formed in the two-dimensional electron 
system (2DES) located 90~nm below the surface. 
The left and right tunnel barriers determine the tunnel coupling of the double dot to 
the drain and source contact, respectively.
The coupling conductance ($G_{c}$) between the dots is controlled 
by the central gate voltage ($V_t$) (see Fig.~\ref{structure}). 
At 4.2~K the carrier density of the 2DES is $1.7 \times 10^{15}$~m$^{-2}$ and 
the electron mobility is 80~m$^{2}$/Vs, 
yielding a mean free path of around 5~$\mu$m which is almost one order of magnitude larger than 
the dots' effective diameters ($\sim 0.6~\mu$m). 
The 2DES is cooled to a bath temperature of 140~mK  
in a $^{3}$He/$^{4}$He dilution refrigerator. 
To couple microwave radiation to the quantum dots, 
we use a Hertzian wire-loop antenna about 1~cm above the dots, 
which is fed with an HP83711A microwave generator~\cite{qin-prb01}.

\begin{figure}[!h] 
\begin{center}
\epsfig{figure=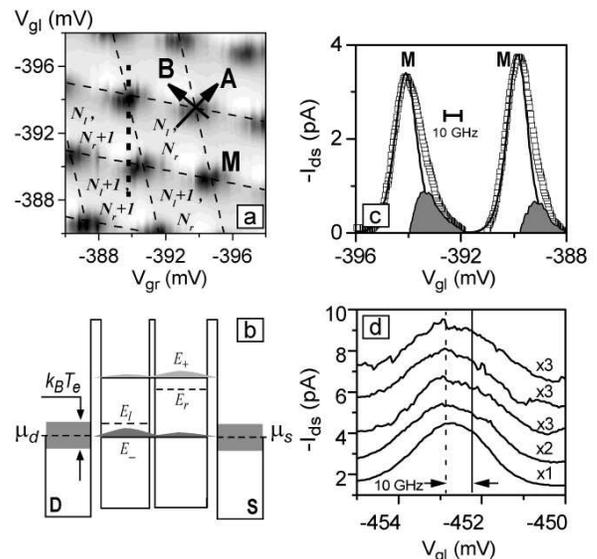,width=8cm,keepaspectratio}
\caption{(a) Charging diagram of drain-source current under weak tunnel coupling ($G_{c} \approx 0.08~e^2/h$) 
~(linear scale, white: $I_{ds} \approx 0$~pA, black: $I_{ds} \geq -4$~pA). 
The charge configuration in the dots is denoted with ($N_l$, $N_r$). 
(b) The level diagram illustrates the resonant tunneling through the bonding molecular state. 
(c) A single trace extracted from (a) at a constant voltage ($V_{gr}$) marked by a vertical dashed line. 
(d) Temperature dependence of the peak asymmetry: 
At bath temperatures of 276, 320, 468, 564, and 634~mK for the curves from bottom to top, respectively. 
} \label{properties}
\end{center}
\end{figure}

The double dot is weakly coupled to the drain and source contacts, 
i.e., the tunneling rates of the outer left and right tunnel barriers are fixed around 
$\Gamma_{l} \approx \Gamma_{r} \sim 100$~MHz
($h\Gamma_l$, $h\Gamma_r \ll k_BT$, $\Delta \epsilon^*$, see below for the excited-state energy $\Delta \epsilon^*$). 
In the measurements,  the drain-source bias $V_{ds}$ is fixed at around $-10~\mu V \ll \Delta \epsilon^*/e$. 
When $V_{ds} < 0$ electrons tunnel through the double dot from drain to source.   
For characterization, charging diagrams are recorded by measuring 
the drain-source current while altering the left ($V_{gl}$) and right ($V_{gr}$) plunger gate voltages. 
We find the effective energy calibration factors $\alpha_{l/r}$ of $6 \times 10^{-2}$ 
for both the left and right gate voltages  
and the addition energies are $351~\mu$eV and $232~\mu$eV for the left and right dot, respectively~\cite{alpha}. 
Both dots show excited states at $\Delta \epsilon^* \approx 120~\mu$eV.
In Fig.~\ref{properties}(a) the charging diagram of the coupled dot is plotted in a linear grayscale representation 
in the weak coupling regime ($G_{c} \approx 0.08~e^2/h$). 
In linear transport, only two ground states participate, e.g., 
the ground state $E_l$ in the left dot and $E_r$ in the right dot, 
as schematically shown in Fig.~\ref{properties}(b)~\cite{simplepicture}. 
The detuning between the two ground states is $E_{lr} = - (E_l-E_r)$. 
The finite tunnel coupling $t$ between these two states induces a bonding ($E_{-}$) 
and an anti-bonding ($E_{+}$) molecular state, as shown in Fig.~\ref{properties}(b). 
The energetic offset between these states is $E_{+}-E_{-} = \sqrt{E_{lr}^2+4t^2}$~\cite{stafford-prl96,stoof-prb96}.   
For weak tunnel coupling and non-zero detuning ($E_l \not = E_r$) it follows that 
a bonding electron is localized in one of the dots (as shown in Fig.~\ref{properties}(b)). 
At zero detuning ($E_l = E_r$), the bonding electron tunnels back and forth between the dots 
at a Rabi frequency of $\Omega_R = t/h$.  
The discrete conductance peaks in Fig.~\ref{properties}(a) are main peaks (M) resulting 
from resonant tunneling through the molecular states with $E_{lr}  = 0$. 

Within the diamond shaped regions (Fig.~\ref{properties}(a)) enclosed by dashed lines 
Coulomb blockade prevails and the number of electrons in the dots is well defined, 
corresponding to the charge configuration ($N_l$, $N_r$). 
In the direction marked by arrow {\bf A} in Fig.~\ref{properties}(a), 
the detuning $E_{lr}$ is kept constant, 
while the average of the ground state energies $\bar E = (E_{l}+E_{r})/2$ is increased 
relative to the chemical potentials ($\mu_{d}$, $\mu_{s}$) of the contacts. 
In direction of {\bf B}, $\bar E$ is fixed but $E_{lr}$ is varied. 
We find that the broadening in both directions is dominated by the  temperature 
but not the intrinsic lifetime of the quantum levels: A signature of weak coupling to the contacts.   
The electron temperature in the contacts~($T_e \approx 170$~mK) 
is obtained simply by measuring the full-width-at-half-maximum~(FWHM) 
of the conductance peak of a single quantum dot~\cite{Beenakker-prb91,foxman-prb93,goldhaber-prl98}. 
In direction of {\bf B}, the FWHM of about $70~\mu$eV is lower than that 
in direction of {\bf A} ($\sim 100~\mu$eV)~\cite{splitting}.  
The level diagram shown in Fig.~\ref{properties}(b) takes into account the 
temperature broadening in the contacts. 

The charging diagram shown in Fig.~\ref{properties}(a) exhibits electron transport by thermally induced phonons. 
To clarify this, a typical single trace extracted from the diagram is shown in Fig.~\ref{properties}(c). 
On the left side of the peak, where $E_l > E_r$, the data is well approximated by $\cosh^{-2}(E_{lr} /2k_{B}T_{e})$. 
This implies that transport at the center of the peak is dominated by 
resonant tunneling through a molecular state with $E_{lr} =0$~(see the solid line in Fig.~\ref{properties}(c)). 
However, on the right side where $E_l < E_r$, an additional contribution to the tunneling current is found, 
which has a maximum~(see the dark region in Fig.~\ref{properties}(c)) about 10~GHz away from the main peak. 
By increasing the bath temperature, we find the position of this maximum to 
shift slightly to the right side, as shown in Fig.~\ref{properties}(d). 
We can exclude both electronic excitations from the drain and 
source contacts ($|eV_{ds}| \approx 10~\mu$eV~$< h \cdot 3$~GHz) and 
intrinsic excitations in the dots ($\Delta \epsilon^* \approx 120~\mu$eV~$ \approx h \cdot 30$~GHz) 
inducing this additional tunneling. 
Furthermore, tunneling through the bonding or anti-bonding molecular state 
with $E_{lr} \not= 0$ is strongly suppressed in this weak coupling regime ($E_{+}-E_{-} \cong E_{lr}$). 
Hence, we can attribute this additional off-resonant tunnel current to be induced by a background of acoustic phonons. 

\begin{figure}[!h]
\begin{center}
\epsfig{figure=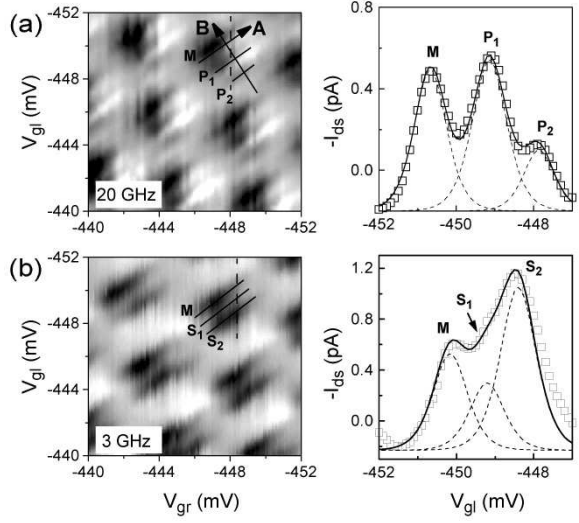,width=8cm,keepaspectratio}
\caption{Charging diagrams of the weakly coupled double dot ($G_c \approx 0.08 e^2/h$) under microwave 
radiation at (a) $20$~GHz and (b) $3$~GHz ~(white: $I_{ds} \approx 0$~pA, black: $I_{ds} \geq -4$~pA). 
The corresponding single traces extracted from the linear grayscale plot 
at a constant $V_{gr}$ (marked by dashed lines in the charging diagrams) show the details. 
} \label{cd-mw}
\end{center} 
\end{figure}

\begin{figure}[!h]
\begin{center}
\epsfig{figure=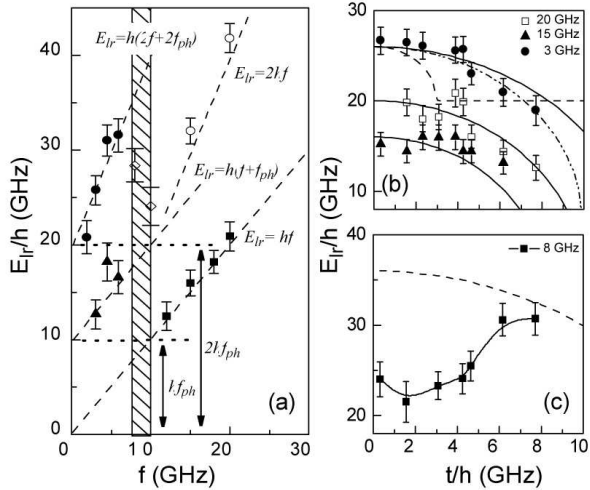,width=8cm,keepaspectratio}
\caption{(a) The detuning at side-peaks P$_1$, P$_2$, S$_1$ and S$_2$ versus microwave frequency 
in the weak tunnel coupling regime ($G_{c} \approx 0.08~e^2/h$). 
$f > 10$~GHz: Solid squares and open circles are for side-peaks P$_1$ and P$_2$, respectively.
$f < 8$~GHz: Solid triangles and solid circles stand for side-peaks S$_1$ and S$_2$, respectively. 
In the hatched region (8~--~10~GHz), the open diamonds correspond to side-peaks S$_2$.
(b) The coupling dependence of the detuning at side-peaks P$_1$ and S$_2$ at 3, 15, and 20~GHz. 
For 15 and 20~GHz, the solid lines correspond to Eq.~[\ref{e_patcoupling-1}]. 
For 3~GHz, the dashed line is from Eq.~[\ref{e_patcoupling-2}], 
the dashed/dotted line is fitted with Eq.~[\ref{e_pn}], and the solid line is based on Eq.~[\ref{e_pnpt}]. 
(c) At 8~GHz, the dashed curve is calculated from Eq.~[\ref{e_pnpt}].
} \label{ft}
\end{center} 
\end{figure}

Application of microwave radiation allows us to perform spectroscopy on single quantum states 
by measuring the direct photo-response~\cite{oosterkamp-nature98,stafford-prl96,stoof-prb96}. 
Fig.~\ref{cd-mw}(a) shows a charging diagram of the weakly coupled dot~($G_c \approx 0.08 e^2/h$) 
under microwave radiation at a frequency of $f = 20$~GHz. 
As seen in the right panel of Fig.~\ref{cd-mw}(a), in addition to the main resonant tunneling peak (M) 
two side-peaks (P$_1$, P$_2$) appear.  
Tunneling at the side-peaks is enabled by the resonance between the microwave photons and 
the energy difference between the molecular states: $E_{+}-E_{-} = nhf \cong E_{lr}$, $n=1, 2, \ldots$.  
Side-peaks P$_1$ and P$_2$ correspond to one-photon ($n=1$) and two-photon ($n=2$) absorption processes, respectively.  
In the right panel of Fig.~\ref{cd-mw}(a), the solid line is an approximation including 
three peaks according to $\cosh^{-2}((E_{lr}-nhf)/2k_{B}T_{e})$, 
with $n = 0$ for the main peak (M) and $n = 1, 2$ for P$_1$ and P$_2$. 
For frequencies above 10~GHz this can be clearly seen in Fig.~\ref{ft}(a) 
where filled squares (open circles) denote the distance of the P$_1$ (P$_2$) side-peaks.

Below 10~GHz, two side-peaks (S$_1$, S$_2$) are resolved from the charging diagrams. 
In Fig.~\ref{cd-mw}(b), the charging diagram at 3~GHz is shown. 
The single trace in the right panel is approximated by  three $\cosh^{-2}(E_{lr} /2k_{B}T_{e})$-shaped peaks. 
The peak S$_1$ is masked between the main peak M and the side-peak S$_2$. 
The existence of S$_1$ is confirmed under microwave radiation at 3, 4.5, and 5.9~GHz.
For the detuning at side-peaks S$_1$ and S$_2$, we find the following relations: 
\begin{eqnarray} 
\label{e_offset-1} 
&&E_{lr}^{S_1} \approx h(f+f_{0}), \\
\label{e_offset-2} 
&&E_{lr}^{S_2} \approx 2h(f+f_{0}),                
\end{eqnarray} 
with a constant offset of $f_{0} \approx 10$~GHz, as shown in Fig.~\ref{ft}(a). 
The slopes of the traces correspond to absorption of one and two photons. 
The offset, however, indicates that transport through the dots involves 
not only pure photon absorption processes, but also another absorption process of $hf_0$ or $2hf_0$. 

We attribute this offset energy quanta in Eqs.~[\ref{e_offset-1},\ref{e_offset-2}] 
to acoustic phonons at $f_{ph} \approx f_0 \approx 10$~GHz.  
As discussed above, we observe thermally induced acoustic phonons 
already without microwave radiation, as shown in Fig.~\ref{properties}(c,d). 
Additionally, the Schottky gates on the heterostructure surface not only define the quantum dots, 
but also form phonon cavities since Al$_x$Ga$_{1-x}$As/GaAs is a piezoelectric material~\cite{fujisawa-science98}.
Electrons in the quantum dots confined only 90~nm below the heterostructure surface thus mainly couple  
to local phonon modes within the cavities. 
Furthermore, the applied microwave field enhances via piezoelectrical coupling the interaction of 
electrons and phonons localized within the cavities. 
Experimentally, strong photon side-peaks shifted by $f_{ph} \approx 10$~GHz are observed 
in the microwave frequency range from 1 to 6~GHz. 
This phonon frequency is determined by the size of the cavities. 
Assuming a sound velocity of about $4500$~m/s in bulk GaAs crystals and 
a phonon frequency of 10~GHz, we find a wavelength of 450~nm. 
This roughly agrees with the dots' diameters as shown in Fig.~\ref{structure}. 
Moreover, this piezoelectrical resonator formed by Schottky gates possesses a large bandwidth, 
due to the limited number of gates in comparison to conventional IDTs~\cite{wixforth-prb89}. 
Hence, exciting this transducer off resonance will also generate phonons 
at 10~GHz as verified experimentally. 
Most importantly, confined phonons generated by the microwave field are 
naturally coherent with the microwave photons. 
Finally, no significant heating by microwaves is observed. 
Indeed, the much stronger side-peaks under radiation (see Fig.~\ref{cd-mw}) 
compared to the weak shoulder without radiation (see Fig.~\ref{properties}(c)) 
confirm that the confined phonon mode at $f_{ph} \approx 10$~GHz in the cavities is enhanced.

The central question now is to verify the coherent superposition of molecular states 
by microwave photons under the excitation of acoustic phonons. 
For this purpose, charging diagrams under radiation are measured upon increasing the tunnel coupling. 
>From these diagrams, the coupling dependence of the detuning of two quantum states, 
at which the two molecular states are in resonance with microwave photons, are obtained. 
As shown in Fig.~\ref{ft}(b), we find that at 15 and 20~GHz the detuning at side-peak P$_1$ 
follows the well-known relation 
\begin{equation}
\label{e_patcoupling-1}
E_{lr} = \sqrt{(hf)^2-4t^2},
\end{equation} 
where it is simply assumed that $t \propto G_{c}$. 
Eq.~[\ref{e_patcoupling-1}] is a signature of a coherently coupled two-level system 
as studied before~\cite{oosterkamp-nature98,stafford-prl96,stoof-prb96}.

Commonly, one expects no coherent superposition of the molecular states 
by single-photon resonance below $f = 8$~GHz, since the thermal fluctuations are comparable to the photon energy. 
Nevertheless, the coupling dependence obtained at 3~GHz turns out to be quite similar to those at 15 and 20~GHz: 
Due to the participation of both photons and phonons the detuning at resonance is larger than without phonons. 
Furthermore, the detuning at resonance decreases as the tunnel coupling is enhanced (see Fig.~\ref{ft} (b)), 
behaving like Eq.~\ref{e_patcoupling-1}. 
Considering the dynamics of electron transport at 3~GHz, the processes of phonon and photon absorption 
between the dots can be coherent as well as sequential. 
The first sequential process to be considered is the following: Phonon absorption occurs within the single dots 
exciting an electron into an intermediate state, while photon absorption couples the dots, i.e. 
\begin{equation}  
\label{e_patcoupling-2}
E_{lr} = \sqrt{(2hf)^2-4t^2}+2hf_{ph},
\end{equation}  
being depicted by the dashed curve in Fig.~\ref{ft}(b) for the side-peaks S$_2$~\cite{whynots1}, 
which is obviously not observed. The other sequential process is given 
by the absorption of photons in the dots but the dots are coupled by phonon absorption. 
The corresponding coupling dependence follows 
\begin{equation}  
\label{e_pn}
E_{lr} = \sqrt{(2hf_{ph})^2-4t^2}+2hf,
\end{equation}  
depicted by the dashed/dotted trace in Fig.~\ref{ft}(b), which is in reasonable agreement with the experimental data. 
On the other hand, assuming coherent photon and phonon generation by the microwave source, 
an electron tunneling coherently between the dots can absorb both photons and phonons. 
The coupling dependence is then written as 
\begin{eqnarray}  
\label{e_pnpt}
E_{lr} = \sqrt{\Big[ 2hf+2hf_{ph} \Big]^2-4t^2},
\end{eqnarray}  
illustrated by the solid curve for $f=3$~GHz in Fig.~\ref{ft}(b). 
Within the error bars also this process can be assumed. 
The coherent absorption of both two-phonons and two-photons by an electron can be viewed as 
a two-quanta absorption process, while the sequential absorption is a four-quanta process, 
which has an even smaller probability. 
The observed strong side-peaks {S$_1$} and {S$_2$} at 3, 4.5, and 5.9~GHz imply the coherent absorption of 
both microwave photons and acoustic phonons. 
This is supported by the fact that 
microwave photons are excited coherently with acoustic phonons via the Schottky gates. 
The superimposed molecular states in the double dot are not demolished by absorbing acoustic phonons,  
which is in agreement with the theoretical prediction by Smirnov {\it et al.}~\cite{smirnov-apl00}. 
Finally, the hatched area in Fig.~\ref{ft} (a), marks a transition region between 8 and 10~GHz. 
An increase of the coupling at 8~GHz enhances the detuning at resonance (see Fig.~\ref{ft} (c)),   
i.e., in this transition region no simple superposition of photon and phonon modes exists.

To summarize we connect two quantum dots of medium size and probe the coherent coupling of the discrete states 
under near zero bias by using microwave spectroscopy. 
The applied radiation coherently excites photons and acoustic phonons via the Schottky gates. 
Apart from conventional photon-assisted tunneling observed above 10~GHz, 
we find a coherent superposition of photon- and phonon-assisted tunneling below 8~GHz. 
We conclude that molecular states in the double dot persist under the excitation of acoustic phonons.

We like to thank W. Zwerger, M. Shayegan and J. P. Kotthaus for helpful discussions. 
This work was funded in part by the Deutsche Forschungsgemeinschaft within 
the project SFB~348, the Bundesministerium f\"ur Forschung und Technologie (BMBF)
within the project Quantenstrukturbauelemente (01BM914), 
and the Defense Advanced Research Projects Agency (DARPA) within the 
Ultrafast Electronics Program (F61775-99-WE015). 
HQ gratefully acknowledges support by the Volkswagen Stiftung.


\newpage

\begin{thebibliography}{11}
\bibitem{alpha} The factors $\alpha_{l}$ and $\alpha_{r}$ are defined as
$\alpha_{l/r}=\delta E_{lr}/e \delta V_{gl/gr}$, 
where $E_{lr}=-(E_{l}-E_{r})$ represents 
the detuning between two quantum states in the left and right dot.
The addition energy is defined as $E_{\Sigma} = e^2/C_{\Sigma} + \delta \epsilon$, 
where $C_{\Sigma}$ and $\delta \epsilon$ represent the total capacitance 
and the level spacing for the single dots.
\bibitem{simplepicture} For simplicity, we adopt a rather simple picture here. 
In fact, $E_{l/r}$ represents the chemical potential of the left/right dot 
associated with the ground state transition $N_{l/r}+1 \leftrightarrow N_{l/r}$ when 
it is aligned with the chemical potential of the drain/source contact.
\bibitem{whynots1} Here we only show the coupling dependence of 
the detuning at side-peak S$_2$ since it has smaller error than that of side-peak {S$_1$}.
\end{thebibliography}
\end{document}